\documentclass[conference]{IEEEtran}

\makeatletter
\let\@ORGmakecaption\@makecaption
\long\def\@makecaption#1#2{\@ORGmakecaption{#1}{#2}\vskip\belowcaptionskip\relax}
\makeatother

\usepackage{listings}

\usepackage{graphicx}
\DeclareGraphicsExtensions{.jpg,.jpeg,.png}

\begin{document}
\title{Bind: a Partitioned Global Workflow\\ Parallel Programming Model}
\author{
\IEEEauthorblockN{Alex Kosenkov\IEEEauthorrefmark{1}, Matthias Troyer\IEEEauthorrefmark{1}\IEEEauthorrefmark{2}}
\IEEEauthorblockA{\IEEEauthorrefmark{1}Theoretische Physik, ETH Zurich, 8093 Zurich, Switzerland}
\IEEEauthorblockA{\IEEEauthorrefmark{2} Quantum Architectures and Computation Group, Microsoft Research, Redmond, WA (USA)}
\\
}

\maketitle


\begin{abstract}
High Performance Computing is notorious for its long and expensive software development cycle. To address this challenge, we present Bind: a "partitioned global workflow" parallel programming model for C++ applications that enables quick prototyping and agile development cycles for high performance computing software targeting heterogeneous distributed many-core architectures. We present applications of Bind to Linear Algebra and MapReduce algorithms alongside with performance results.
\end{abstract}

\IEEEpeerreviewmaketitle

\section{Introduction}
At the moment High Performance Computing has a substantial problem, the cost of the software development cycle. It is expensive to develop an application that efficiently utilises available resources, whether it is a cluster with each node equipped with several GPUs or a cloud/cluster mixed deployment. And things only get worse over time as for every change in the computing infrastructure, whether it is a transition to a new CPU family or the addition of new accelerators one has to invest time to tune, optimise and ensure the stability of an application. Sometimes it is even easier to write whole applications from scratch for a particular technology stack.

One of the reasons of such state of things is the abstraction level and complexity of the current HPC software stack. In order to produce an efficient scalable application one has to micro-manage several aspects of code execution.

First, there is a need for a multi-process execution orchestration:
\begin{itemize}
\item{Each process has to be explicitly informed about what should it do and how.}
\item{A process has to know ``when" and ``what" information to send to and receive from other processes.}
\item{Finally, one has to define how this communication is set up and optimize the computation/communication overlap.}
\end{itemize}

Second, threaded execution has to be implemented for each process to efficiently utilise the node's CPU cores:
\begin{itemize}
\item{The process' local tasks and data parallel regions that could benefit from threading should be identified and instrumented.}
\item{Synchronisation between computational and communication threads should be carefully engineered.}
\end{itemize}

Finally, one might want to use accelerators, such as GPGPUs to offload some parts of the computation:
\begin{itemize}
\item{One additionally has to keep track not only of local and remote memory but also to transfer memory between CPU and GPU within the same node or between different nodes.}
\item{The synchronisation between CPU, accelerator and communication becomes even more challenging.}
\end{itemize}

Everything mentioned above does not eliminate the necessity of optimised computational algorithms for the specific task at hands. 

Here we propose a solution to many of these challenges by introducing Bind, a "partitioned global workflow" parallel programming model that deals with them in several ways:

\begin{itemize}
\item{It is based on the classical sequential code design.}
\item{Threading is performed automatically.}
\item{Data transfer is implicit.}
\item{Program execution is reproducible.}
\item{Race conditions are avoided by design.}
\item{Portable and easy to install by being a header-only library based upon MPI}
\end{itemize}

The points above immediately translate into accelerated and simplified code development by keeping the user code agnostic of the low-level aspects of the parallel application execution.

\section{Key Principles}
\subsection{Global Workflow}
We start by defining what we mean by ``global workflow" is and why it is important. The foundation of Bind's programming model is built on the notion of dataflow that is a set of parallel computational stages connected by data transfers. But unlike some dataflow programming languages \cite{ackerman1982data}, Bind does not require users to reformulate the code in terms of pipelines and instead relies on the directed acyclic graph (DAG) of operations that is extracted dynamically from the user's serial code execution flow (workflow).

Once the code is decomposed into a set of computational stages one can easily execute them in parallel with respect to data constraints (i.e. computational dependencies). And more importantly, move some of the stages onto other nodes, hence enabling distributed memory parallelisation.

For a group of processes to be used this way, each of them has to have information about which actions each of them has to perform on what data. And in order for such a system to scale while remaining simple and robust one has to avoid centralisation of this information on any particular node. For this reason, Bind's actual workflow is "global" that is produceable independently by any process in a group by means of parsing the sequential user code execution flow. This makes a sequential code base not just a benefit but a requirement since execution divergence would lead to inconsistency between processes' local representations of the global operational DAG.

\subsection{Multi-version concurrency}
The construction of the DAG used as a basis for the parallel execution is by no means a trivial task. Bind uses function decorators and multi-version concurrency control \cite{bernstein1983multiversion}. By employing function decorators Bind is able to inspect a function's prototype at compile time. In particular it is able to detect which arguments are going to be modified and which arguments are going to remain constant. This information is then recorded in terms of argument object's ``states". Specifically, if a function takes in an object without a ``const" modifier Bind assumes that the object will change and this is accounted for by creating a new ``version" of this object, marking the function call as a generator for this version. Thus, for every operation in a program it is known which version of which object is needed and which operations are responsible for their input generation. This composes all operations into a transactional DAG.

The concept of object's versioning plays a key role here and in the overall programming model.
Besides serving to construct the operational DAG it enables an additional degree of parallelism by allowing multiple versions of the same object to be used simultaneously. This means that operations depending on the newer state of the object need not wait until all of the older operations will finish. For example an MPI transfer will not block the next operations.

Additionally, by its nature the multi-version concurrency control guarantees that race conditions will never occur since a version of the object is immutable. This, coupled with the deterministic DAG construction, makes a program execution reproducible and free of race conditions.

\begin{figure}[!t]
\centering
\includegraphics[width=3.2in]{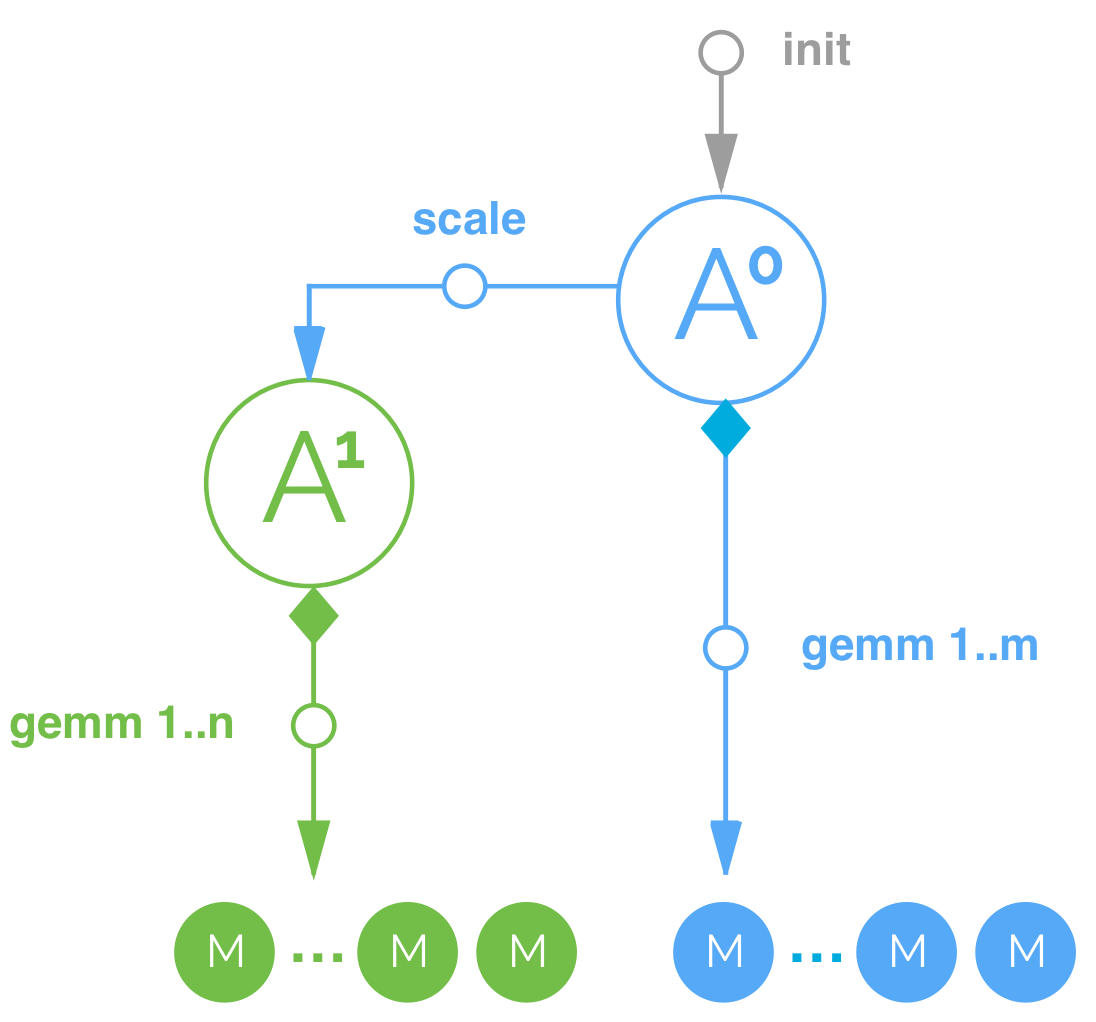}
\caption{Transactional DAG example. Two sets of matrix multiplications (of size $n$ and $m$) depend on two different states of matrix A. Keeping both states (before and after scaling) allows to perform $n+m$ operations in parallel.}
\label{fig:dag}
\end{figure}

\subsection{Partitioning}
While the resulting DAG contains all of the necessary information to be executed in a fully distributed way automatically, Bind still leaves the declarative partitioning of the sequential code to the user\footnote{Although (semi-)automatic scheduling is present in Ambient (a parallel framework based upon the Bind model).}. In other words, the programmer has to declare on which node a set of operations has to be executed by means of scope guards. Once it is done, Bind takes care of the actual data transfers.

The reason is that optimal scheduling of the DAG across many nodes is a hard optimisation problem. While heuristical solutions \cite{gerasoulis1992comparison} might be good enough for shared-memory architectures the penalty for a suboptimal distribution is too high in case of the distributed systems due to high communication costs. It gets even more problematic if the whole DAG is not known before execution, since an effective solution in one part of the DAG might lead to inferior scheduling for the future parts of the DAG.

\section{Performance Implications}
As was mentioned in the introduction, Bind can substantially reduce development time and cost by providing safety (which reduces debugging efforts) and abstraction from the low level details of the parallelisation (such as threading and communication). However, by virtue of its name, performance is crucial to HPC. And it turns out that the concept of the transactional DAG also provides performance advantages:
\begin{itemize}
\item{\bf Zero-copy:}
due to the multi-versioned nature of every object that is a part of the transactional DAG, there is no need to perform any copies of data. Instead it is sufficient just to point a pointer to the right revision of the object's data. This happens completely transparently for the user without a need to change classical pass-by-value semantics.
\item{\bf Memory differentiation:}
as a side effect of having a transactional DAG before its actual execution, Bind can choose which type of memory the object should be allocated in depending on the object's lifetime and behaviour. E.g. it can get a pinned memory if it is going to be frequently transferred or bulk memory if the lifetime of the object is short. This can be crucial for applications sensitive to memory allocation strategies or memory-bound applications.

\item{\bf Memory locality and defragmentation:}
since objects get frequently reallocated due to versioning, one automatically gains improved memory locality for the objects that are used jointly, thus improving TLB cache hit rate \cite{saavedra1995measuring}. This also helps avoid memory fragmentation issues in the long-running applications with persistent objects, thus improving allocations system time and memory capacity of the host.
\item{\bf Computation/communication overlap:}
all of the implicit communication resulting from user-defined partitioning is performed completely asynchronously, thus allowing to efficiently overlap computations and communications. The intrinsic model's ability to retain multiple versions of the same data greatly facilitates this process since the computation becomes completely decoupled from the state of the memory transfers.
\item{\bf Implicit collectives:}
inevitably there are cases when all of the processes would need to operate on the same dataset thus requiring a collective communication. Bind's model infers such cases from the globally available DAG of operations and their corresponding locations. By doing that it not only covers the functionality of global collectives but automatically infers "partial" collectives \cite{hoefler2009sparse} (e.g. when only half of the nodes would need some dataset). It does so by dynamically constructing a binary tree from the queue of the communications involving the same object across multiple nodes.

\item{\bf Lockless execution:}
finally, another performance advantage of the transactional nature is the absence of any internal locks or synchronisation primitives. This directly stems from the fact that the transaction can be either completed or not, thus keeping all of its dependencies in the waiting queue. Thus avoiding mutex handling and related memory/cache coherence CPU cycles.
\end{itemize}

The negative aspects of our model are a bigger memory requirement and an overhead of the run-time DAG construction.
The bigger memory requirement stems from multi-versioning meaning that an object may occupy more memory if it exists in multiple versions simultaneously, which is usually related to the degree of the parallelism the model exposes (with smart memory reusage to mitigate the overhead when possible).

On the other hand, run-time construction of the DAG (despite being carefully optimised by means of template meta-programming) can still incur a critical disadvantage depending upon the computational cost of a single operation.

\section{Benchmarks}
In order to demonstrate the merits of our programming model, this section will cover two examples: matrix multiplication and sorting of integers using the MapReduce pattern.

\subsection{Matrix Multiplication}

\begin{figure}[!t]
\centering
\includegraphics[width=3.5in]{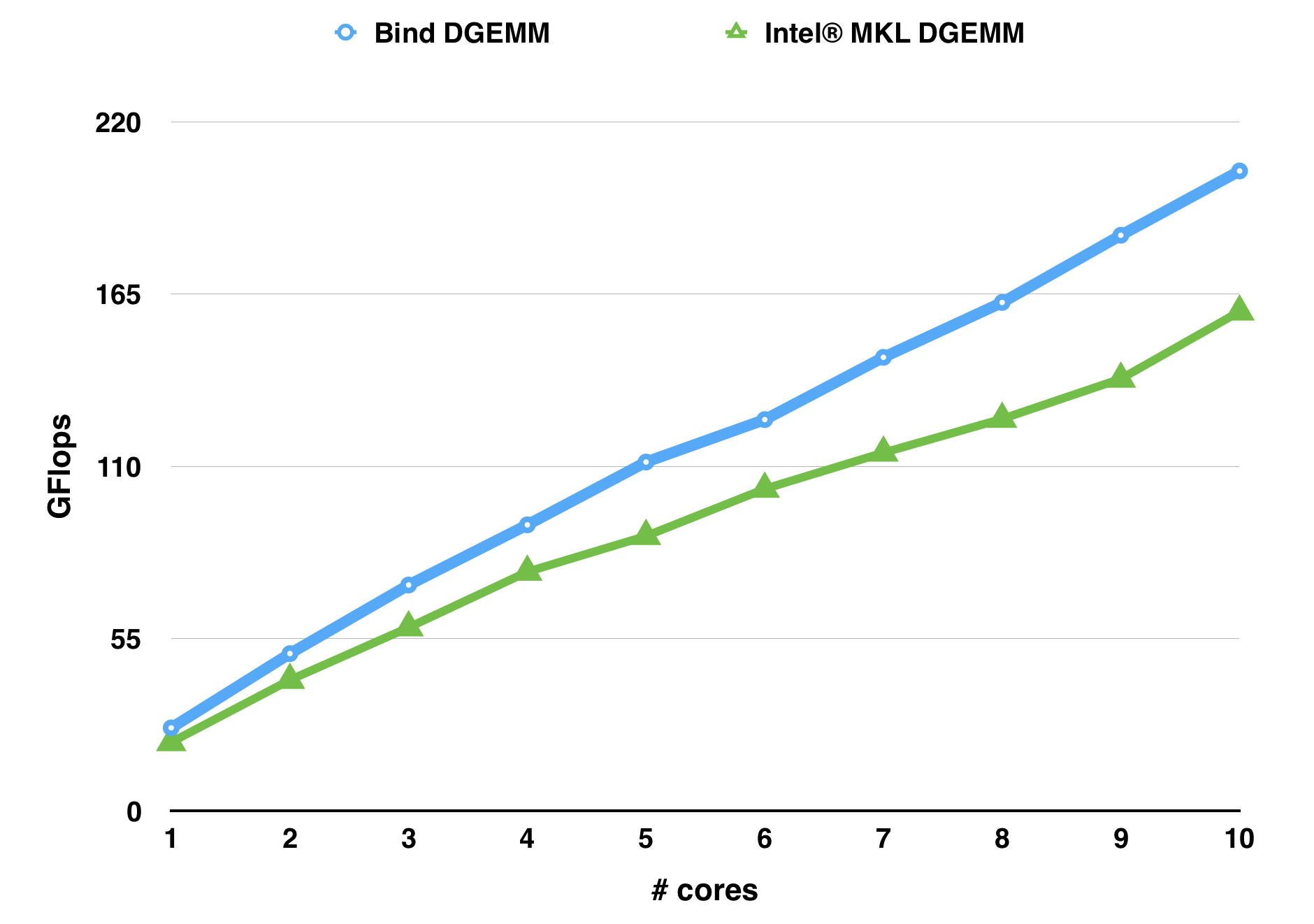}
\caption{Strassen matrix multiplication performance.}
\label{fig:strassen_perf}
\end{figure}

For matrix multiplication we consider two algorithms: a shared-memory implementation of Strassen's algorithm \cite{strassen1969gaussian} and distributed classical matrix multiplication algorithm with logarithmic reduction.

The performance results were obtained on CSCS Monch cluster, which consists of two-socket 10 cores Ivy Bridge (Intel\textsuperscript{\textregistered} Xeon\textsuperscript{\textregistered} EP E5-2660 v2 @2.2 GHz with 32 GB or RAM) nodes connected into a fat tree topology using Infiniband FDR interconnect.

Figure \ref{fig:strassen_perf} shows a performance comparison between Intel\textsuperscript{\textregistered} MKL's DGEMM and Bind's Strassen matrix multiplication for a square matrix of 8192 by 8192 elements. The latter uses matrices stored as collections of tiles where each tile denotes a rectangular block of its original matrix and is stored contiguously in memory. The algorithm is executed recursively on the tiled matrices and their submatrices until the size of a submatrix hits a single tile; then the operation would be dispatched to the sequential Intel\textsuperscript{\textregistered} MKL's DGEMM call. The DAG yielded by these series of recursive calls is then executed in parallel using Bind's execution engine.

As one can see, on average the MKL's parallel DGEMM requires 25 percents more time than the Strassen's algorithm (with the downside of the Strassen's algorithm requiring more memory).

\begin{figure}[!t]
\centering
\includegraphics[width=3.5in]{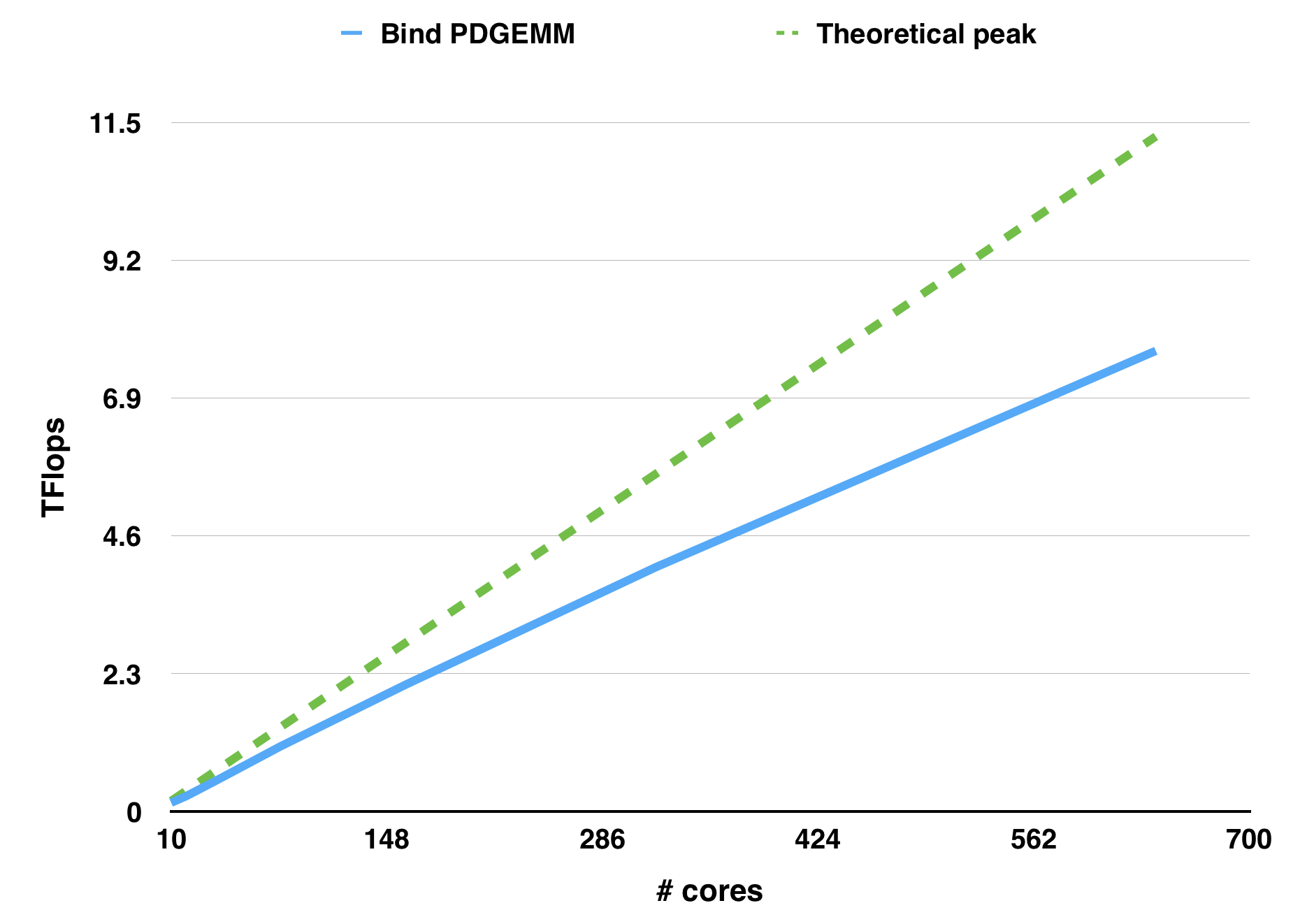}
\caption{Distributed matrix multiplication performance.}
\label{fig:pdgemm_perf}
\end{figure}

\begin{figure}[!t]
\centering
\includegraphics[width=3.5in]{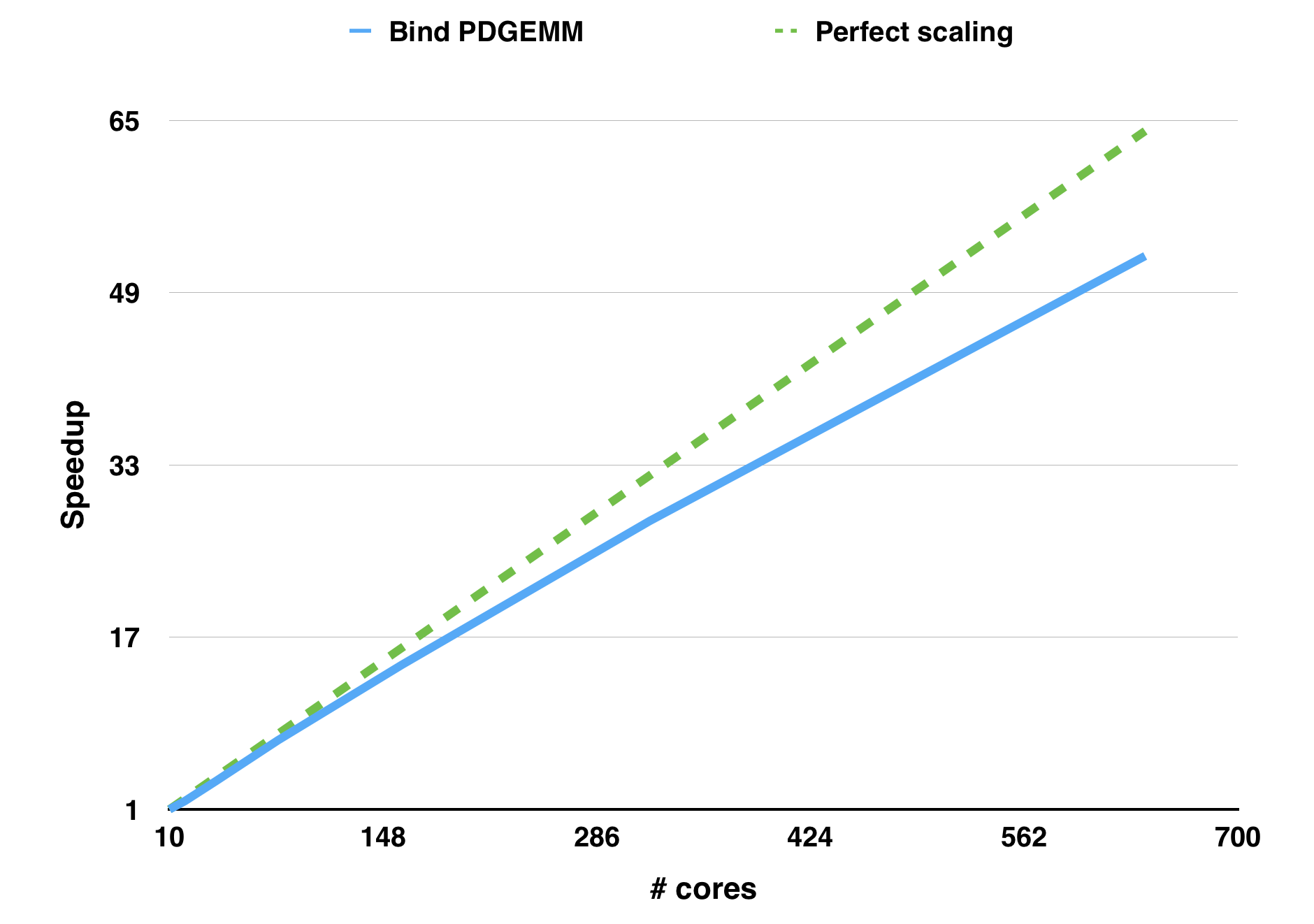}
\caption{Scaling of the distributed matrix multiplication.}
\label{fig:pdgemm_scaling}
\end{figure}

Figure \ref{fig:pdgemm_perf}, in turn, shows the performance of the distributed classical DGEMM relative to the cumulative theoretical peak performance. Since numerical stability is critical for the large scale matrix operations, this algorithm employs logarithmic reduction where any block of output matrix accumulates all of its updates by means of a binary tree reduction.

As one can see, for the given square matrix of 32768 by 32768 elements, 640 cores (64 nodes employing only one socket each) yield about 70\% of the cumulative theoretical peak performance with an algorithm that just uses 18 lines of code (see Listing \ref{lst:pdgemm_code}).

\begin{lstlisting}[
    label={lst:pdgemm_code},
    caption={Distributed matrix multiplication with logarithmic reduction},
    morekeywords={bind,node,sync,gemm},
    belowcaptionskip=1em, aboveskip=1em, belowskip=1em,
    frame=tb,
    basicstyle=\footnotesize\ttfamily
]
for(int ii = 0; ii < a.mt / NP; ii++)
for(int kk = 0; kk < b.nt / NQ; kk++){
   for(int i = ii*NP; i < (ii+1)*NP; i++)
   for(int k = kk*NQ; k < (kk+1)*NQ; k++){
      std::vector<matrix> r(a.nt, c.tile(i,k));
      for(int j = 0; j < a.nt; j++){
         bind::node p((i%NP)*NQ + j%NQ);
         gemm(a.tile(i,j), b.tile(j,k),
                r[(a.nt-k+j) % a.nt]);
      }
      for(int s = 1; s < a.nt; s *= 2)
      for(int w = s; w < a.nt; w += s*2){
         bind::node p((i%NP)*NQ+((k+w-s)%a.nt)%NQ);
         r[w-s] += r[w];
      }
      c.tile(i,k) = r[0];
   }
   bind::sync();
}
\end{lstlisting}

\subsection{Sorting}

In ``Big Data'' applications and data analytics, sorting of large distributed datasets is a common task, which is often performed using Hadoop\textsuperscript{\textregistered} \cite{lam2010hadoop} or Spark{\tiny\texttrademark} \cite{gopalani2015comparing} employing the MapReduce parallel programming model for large scale cloud data processing.

Here, we measure the performance of a trivial implementation of a MapReduce engine using Bind, which can perform map, reduce, combine and implicit shuffle operations. See the integer sorting implementation as shown in Listing \ref{lst:bind_sort_code}.

\begin{figure}[!t]
\centering
\includegraphics[width=3.5in]{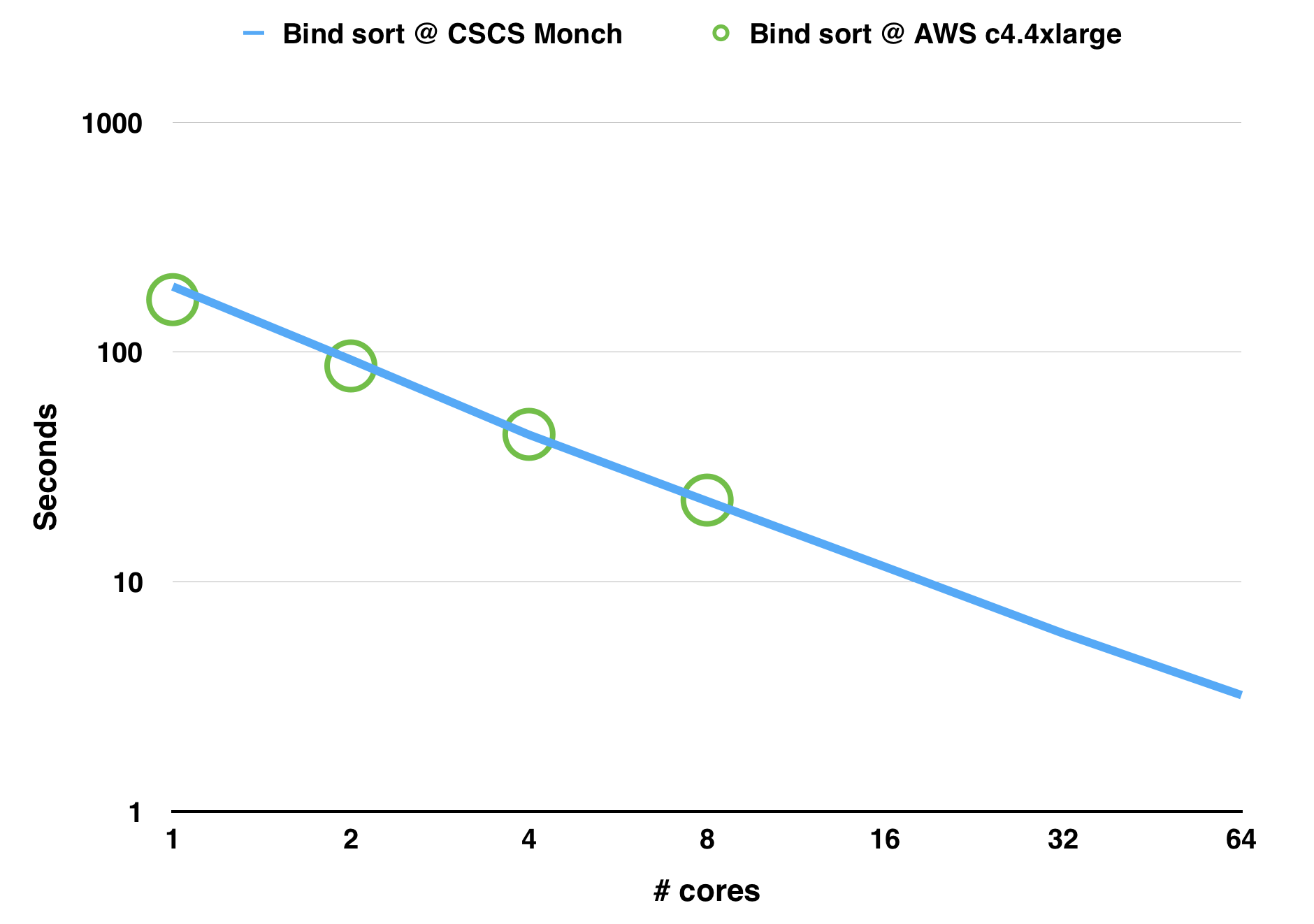}
\caption{Bind sorting performance on 1 billion integers.}
\label{fig:bind_sort}
\end{figure}

Similarly to the real world problems, the test-case was to sort one billion uniformly random integers which is roughly comparable to the amount of active Facebook users.
The test case was benchmarked using a single core per node on the Monch cluster in order to stress the interconnect sensitivity of the resulting application and thus bring it closer to a cloud-like environment. As one can see from the Fig. \ref{fig:bind_sort}, the benchmark demonstrated perfect scalability on up to 64 nodes.

\begin{lstlisting}[
    label={lst:bind_sort_code},
    caption={Sorting integers using Bind's MapReduce},
    morekeywords={map,reduce,KVPairs},
    belowcaptionskip=1em, aboveskip=1em, belowskip=1em,
    breaklines, breakatwhitespace=true,
    frame=tb,
    basicstyle=\footnotesize\ttfamily
]
KVPairs<int, doc_type>(local_map)
.map([](int k, std::vector<doc_type>& docs)
  -> std::vector<std::pair<key_type, value_type>>
{
   std::vector<std::pair<int, value_type>> res;
   for(auto doc : docs) for(auto v : doc){
      key_type bucket = v >> (31 - LOG_BINS);
      res.push_back({ bucket, v });
   }
   return res;
})
.reduce([](key_type k, std::vector<value_type>& vs)
  -> std::vector<std::pair<key_type, value_type>>
{
   std::vector<std::pair<int, value_type>> res;
   std::sort(vs.begin(), vs.end());
   for(auto v : vs) res.push_back({ k, v });
   return res;
});
\end{lstlisting}

\begin{lstlisting}[
    label={lst:spark_sort_code},
    caption={Sorting integers using Apache Spark{\tiny\texttrademark}},
    morekeywords={map,parallelize,sortByKey},
    belowcaptionskip=1em, aboveskip=1em, belowskip=1em,
    frame=tb,
    basicstyle=\footnotesize\ttfamily
]
sortedNumbers = sc.parallelize(range(1, n+1), np) \
   .map(lambda x: (randint(1,2147483647), 1)) \
   .sortByKey()
\end{lstlisting}

In order to get a feeling on how the performance of this MapReduce would compare to the status quo in Big Data, the test case was then moved into the AWS cloud with significantly smaller problem size (8 millions integers) and compared to a similar test-case using Apache Spark{\tiny\texttrademark} (see Listing \ref{lst:spark_sort_code}). The performance data was obtained in a slightly heterogeneous environment with nodes mainly using Intel\textsuperscript{\textregistered} Xeon\textsuperscript{\textregistered} E5-2670 v2 @2.50GHz and E5-2676 v3 @2.40GHz.

\begin{figure}[!t]
\centering
\includegraphics[width=3.5in]{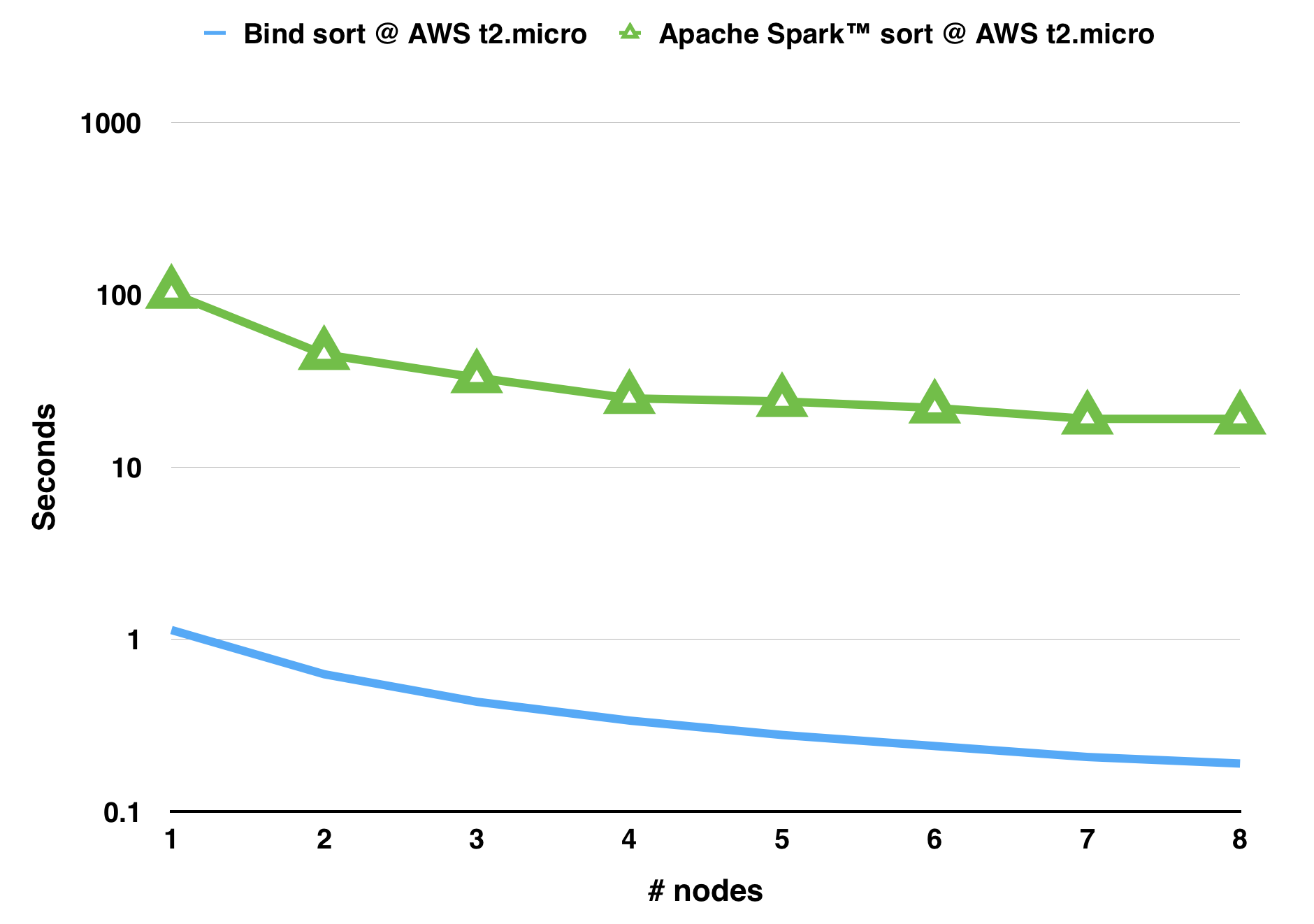}
\caption{Bind and Apache Spark{\tiny\texttrademark} performance on 8 mln integers.}
\label{fig:bind_spark}
\end{figure}

As one can see from the Fig. \ref{fig:bind_spark} while the scalability is quite similar, the absolute performance difference is close to 100 times.

\section{Conclusion}

This paper has given an overview of the key concepts of the partitioned global workflow parallel programming model and its implementation in ``Bind''. The proposed approach based upon extraction of the transactional DAG was shown not only to enable high-productivity software development cycles but also to deliver high performance as was illustrated with several synthetic benchmarks. In particular it was shown that the applications based upon this model are able to achieve close to native hand-crafted MPI based applications performance and scalability on many-core distributed systems both in cloud and cluster environments. More over, it was demonstrated that the performance of Bind based applications can exceed those of the conventional widely used solutions, i.e. at Linear Algebra algorithms or Big Data analysis. Such performance and an abstraction level that comes with this programming model might be as well a key to the whole range of future high performance computing applications that would be easily maintainable and adaptable to potentially disruptive conceptual shifts in computing.

\newpage
Matrix-multiplication examples in this paper were composed by means of Ambient, a parallel library based upon the Bind model.
The source codes of both frameworks can be obtained on the GitHub under Boost license.

\begin{lstlisting}[
    title={Appendix: Strassen matrix multiplication source code},
    belowcaptionskip=1em, aboveskip=1em, belowskip=1em,
    frame=tb,
    float=*,
    basicstyle=\footnotesize\ttfamily
]

    template<class MatrixA, class MatrixB, class Matrix, int IB>
    void gemm_strassen(tiles<MatrixA, IB>&& a, tiles<MatrixB, IB>&& b, tiles<Matrix, IB>&& c){
        size_t n  = c.cols/2;
        size_t nt = c.nt/2;
        if(nt){
            tiles<matrix<value_type>, IB> m1(n, n);
            tiles<matrix<value_type>, IB> m2(n, n);
            tiles<matrix<value_type>, IB> m3(n, n);
            tiles<matrix<value_type>, IB> m4(n, n);
            tiles<matrix<value_type>, IB> m5(n, n);
            tiles<matrix<value_type>, IB> d(n*2, n*2);
            tiles<matrix<value_type>, IB> e(n*2, n*2);

            d = a; m4 = a.subset(0, 0, nt, nt);
            e = b; m5 = b.subset(0, 0, nt, nt);

            d.subset(0,  0,  nt, nt) += a.subset(0,  nt, nt, nt);
            d.subset(0,  nt, nt, nt) -= a.subset(nt, nt, nt, nt);
            d.subset(nt, nt, nt, nt) += a.subset(nt, 0,  nt, nt);
            d.subset(nt, 0,  nt, nt) -= a.subset(0,  0,  nt, nt);

            e.subset(0,  0,  nt, nt) += b.subset(0,  nt, nt, nt);
            e.subset(0,  nt, nt, nt) -= b.subset(nt, nt, nt, nt);
            e.subset(nt, nt, nt, nt) += b.subset(nt, 0,  nt, nt);
            e.subset(nt, 0,  nt, nt) -= b.subset(0,  0,  nt, nt);

            m4 += a.subset(nt, nt, nt, nt);
            m5 += b.subset(nt, nt, nt, nt);
           
            gemm_strassen(d.subset(0,  nt, nt, nt),
                          e.subset(nt, nt, nt, nt),
                          c.subset(0,  0,  nt, nt));
            gemm_strassen(a.subset(0,  0,  nt, nt),
                          e.subset(0,  nt, nt, nt),
                          c.subset(0,  nt, nt, nt));
            gemm_strassen(d.subset(nt, nt, nt, nt),
                          b.subset(0,  0,  nt, nt),
                          c.subset(nt, 0,  nt, nt));
            gemm_strassen(d.subset(nt, 0,  nt, nt),
                          e.subset(0,  0,  nt, nt),
                          c.subset(nt, nt, nt, nt));

            gemm_strassen(std::move(m4), std::move(m5),
                          std::move(m1));
            gemm_strassen(d.subset(0,  0,  nt, nt),
                          b.subset(nt, nt, nt, nt),
                          std::move(m2));
            gemm_strassen(a.subset(nt, nt, nt, nt),
                          e.subset(nt, 0,  nt, nt),
                          std::move(m3));

            c.subset(nt, nt, nt, nt) += c.subset(0,  nt, nt, nt);
            c.subset(nt, nt, nt, nt) -= c.subset(nt, 0,  nt, nt);

            c.subset(0,  0,  nt, nt) += m1;
            c.subset(nt, nt, nt, nt) += m1;
            c.subset(0,  nt, nt, nt) += m2;
            c.subset(0,  0,  nt, nt) -= m2;
            c.subset(0,  0,  nt, nt) += m3;
            c.subset(nt, 0,  nt, nt) += m3;
        }else{
            gemm(a[0], b[0], c[0]);
        }
    }
\end{lstlisting}
\end{document}